\documentclass[preprint,compress]{elsarticle}

\usepackage{bm}
\usepackage{amsmath}

\usepackage[dvips]{color} % For blue in-text comments and additions
\begin{document}
%%%%%%%%%%%%%%%%%%%%%% Title page %%%%%%%%%%%%%%%%%%%%%%%%%%%%%%

\title{
Anomalous-hydrodynamic analysis of 
charge-dependent \\
elliptic flow in heavy-ion collisions
}

\author{Masaru~Hongo}
\ead{masaru.hongo@riken.jp}
\address{
iTHES Research Group, RIKEN, Wako 351-0198,
 Japan
}

\author{Yuji~Hirono}
\ead{yhirono@bnl.gov}
\address{
Department of Physics, Brookhaven National Laboratory, Upton, New York 11973-5000
}

\author{Tetsufumi~Hirano}
\ead{hirano@sophia.ac.jp}
\address{
Department of Physics, Sophia University, Tokyo 102-8554, Japan
}

\date{\today}

\begin{abstract}
Anomalous hydrodynamics is a low-energy effective theory that captures 
effects of quantum anomalies. 
We develop a numerical code of anomalous hydrodynamics and
apply it to dynamics of heavy-ion collisions, 
where anomalous transports are expected to occur. 
This is the first attempt to perform fully non-linear numerical simulations 
of anomalous hydrodynamics.
We discuss implications of the simulations for possible experimental
 observations of anomalous transport effects. 
From analyses of the charge-dependent elliptic flow parameters 
($v_2^\pm$) as a function of the net charge asymmetry $A_\pm$, 
we find that the linear dependence of $\Delta v_2^\pm
 \equiv v_2^- - v_2^+$
 on the net charge asymmetry $A_\pm$
 cannot be regarded as a robust signal of anomalous transports,
 contrary to previous studies.
We, however, find that the intercept $\Delta v_2^\pm(A_\pm=0)$ is sensitive to anomalous transport effects.
\end{abstract}


\maketitle

\emph{Introduction.}---
Macroscopic transport phenomena induced by the triangle anomalies 
have recently attracted much attention.  
One such example is the chiral magnetic(separation) effect (CME/CSE), 
which states that a dissipationless vector(axial) current is generated along a
magnetic field  \cite{Vilenkin:1980fu,Fukushima:2008xe,
Son:2004tq,Metlitski:2005pr}.
This kind of anomalous transport phenomena 
is expected to occur in a variety of materials such as
the quark-gluon plasma (QGP), 
interiors of  neutron stars \cite{Charbonneau:2009ax}, 
or Weyl semimetals \cite{Zyuzin:2012tv,Basar:2013iaa}.
The existence of anomalous transport effects is shown from 
perturbation theory, 
hydrodynamics \cite{Son:2009tf}, and kinetic theory
\cite{Gao:2012ix,Son:2012wh,Stephanov:2012ki,Son:2012zy,Chen:2012ca}.

The quark-gluon plasma (QGP), 
which consists of deconfined quarks and gluons, 
is considered to be a good place to test 
anomalous transport effects. 
The QGP is experimentally produced in heavy-ion collisions 
at Relativistic Heavy Ion Collider (RHIC) in BNL, 
and Large Hadron Collider (LHC) in CERN. 
Through the analysis of the collision experiments, 
the QGP is found to behave like an almost perfect fluid
\cite{Adams:2005dq,Adcox:2004mh,Romatschke:2007mq,Song:2010mg}. 
Since two charged nuclei collide at very high energy, 
extremely strong magnetic fields are created in off-central collisions, 
along which vector and axial currents flow
due to CME/CSE \cite{Vilenkin:1980fu,Fukushima:2008xe,
Son:2004tq,Metlitski:2005pr}.
Recently, it has been theoretically discovered that 
interplay between
CME and CSE results in a novel type of the gapless excitation, 
called the chiral magnetic wave (CMW) 
\cite{Kharzeev:2010gd}. 
The CMW is a charge propagating wave along magnetic fields. 
It brings a charge quadrupole deformation in the QGP, 
which could be used as an experimental signal of anomalous
transport effects \cite{Burnier:2011bf,Burnier:2012ae}.
The results of linearized calculations in a uniform static temperature 
seem to be consistent with experimental data
\cite{Wang:2012qs,Ke:2012qb}.  
However, linearization procedure  is by no means justified since 
it lacks the dynamics of the QGP fluid 
which is crucially important in order to compare 
theoretical calculations with experimental data.
Thus, in order to assess the contribution from the anomalous transport
in the QGP, 
fully non-linear 
numerical simulations of anomalous hydrodynamics are indispensable, 
which have not been performed yet.

In this Letter, we perform the first numerical simulation 
of anomalous hydrodynamics in (3+1) dimensions. 
This enables us to 
estimate the effect of anomalous transport quantitatively. 
We apply anomalous hydrodynamics to
the space-time evolution of the QGP created in heavy-ion collisions and
 demonstrate how the chiral magnetic wave propagates in an expanding plasma. 
We claim that the slope parameter $r_e$, which is
recently proposed 
as a signal of the CMW \cite{Burnier:2011bf,Burnier:2012ae}, 
is not a robust
 signal of anomalous transport.
Rather, the charge-dependent elliptic flow $v_2^\pm$ at zero charge
asymmetry turns out to be sensitive to the anomalous transport effects
in a robust way. 

\emph{Hydrodynamics with the triangle anomaly.---}
Here we briefly discuss
formulation of anomalous hydrodynamics \cite{Son:2009tf,Kalaydzhyan:2011vx}.
We consider a system with one $U(1)_{\rm V}$ vector current 
$j^\mu$,
and one $U(1)_{\rm A}$ axial vector current 
$j_5^\mu$. 
The axial current is not conserved because of the axial anomaly.
Hydrodynamic equations, therefore, represent (non-)conservation laws of
energy, momentum, vector and axial currents;
\begin{eqnarray}
 \partial_\mu T^{\mu \nu} &=& eF^{\nu\lambda}j_\lambda , \label{eqn-emcon}\\
 \partial_\mu j^\mu &=&  0 , \label{eqn-vccon} \\
 \partial_\mu j_5^\mu &=&  -CE^\mu B_\mu, \label{eqn-accon} 
\end{eqnarray}
where $T^{\mu \nu}$ is the energy-momentum tensor of the fluid, 
$e$ the elementary charge, $F^{\mu\nu}$ 
the electromagnetic field strength tensor, 
$E^\mu \equiv F^{\mu \nu} u_\nu,
 \ B^\mu \equiv \tilde{F}^{\mu \nu}u_\nu = \frac{1}{2}
 \epsilon^{\mu\nu\alpha\beta}u_\nu F_{\alpha \beta}$ the 
electric and magnetic fields, 
and 
$C$ the anomaly coefficient.
We here treat electromagnetic fields as external fields.
The right hand sides of Eqs.~(\ref{eqn-emcon}) and (\ref{eqn-accon}) represent 
contributions of the Lorentz force and the triangle anomaly, respectively.

For a perfect fluid, we can express the energy-momentum tensor using the 
four-velocity $u^\mu=\gamma(1, \vec{v})$ as
\begin{eqnarray}
 T^{\mu \nu} &=& (\varepsilon + P)u^\mu u^\nu - Pg^{\mu \nu },
\end{eqnarray}
where $\varepsilon$ is the energy density, $P$ the pressure,
and $g^{\mu \nu} = \mathrm{diag}(1, -1, -1, -1)$ the metric tensor. 
Although vector and axial currents are non-dissipative in a perfect
fluid, 
they should have space-like components $\nu^\mu$ and $\nu_5^\mu$
in the presence of the triangle anomaly:
\begin{eqnarray}
 j^\mu &=& nu^\mu + \nu^\mu, \\
 j_5 ^\mu &=& n_5 u^\mu + \nu_5^\mu,
\end{eqnarray}
where $n=n_{\rm R} + n_{\rm L}$ denotes the charge density, $n_5 = n_{\rm R} -
n_{\rm L}$ the 
axial charge density, and $n_{\rm R,L}$ the charge density of particles
with right/left-handed chirality.

Corrections from the triangle anomaly, $\nu^\mu$ and $\nu_5^\mu$, 
include novel transport phenomena 
known as the chiral magnetic effect, the chiral separation effect,
and the chiral vortical effect. 
The constitutive equations in the presence of anomaly
can be written in terms of the magnetic field $B^\mu$ and the four-vorticity 
 $\omega^\mu 
= \frac{1}{2}\epsilon^{\mu \alpha\beta \gamma}u_\alpha \partial_\beta u_\gamma$
as
\begin{eqnarray}
 \nu^\mu &=& \kappa_B B^\mu + \kappa_\omega \omega^\mu, \\
 \nu_5^\mu &=&  \xi_B B^\mu + \xi_\omega \omega^\mu ,
\end{eqnarray}
where the transport coefficients are 
given by \cite{Son:2009tf} 
\begin{eqnarray}
 e\kappa_B &=& 
C \mu_5 \left( 1 - \frac{\mu_5 n_5}{\varepsilon + p}\right) , \quad
 e^2 \kappa_\omega 
 = 2C\mu\mu_5\left(1-\frac{\mu n}{\varepsilon +	p}\right) , \nonumber \\
\label{eqn-cmc}
\\
 e\xi_B &=& C \mu \left( 1 - \frac{\mu n}{\varepsilon + p}  \right) , 
\quad
 e^2 \xi_\omega 
 = C\mu^2 \left( 1 - \frac{2\mu_5 n_5}{\varepsilon + p}\right). \nonumber \\
\label{eqn-csc}
\end{eqnarray}
Here, $\mu = \mu_{\rm R} + \mu_{\rm L}$ and $\mu_5 = \mu_{\rm R} -
\mu_{\rm L}$ are the chemical
potentials for the vector and axial charges, respectively.

In this Letter, we adopt the following constitutive equations, 
\begin{eqnarray}
 j^\mu = nu^\mu + \kappa_B B^\mu, \\
 j_5^\mu = n_5 u^\mu + \xi_B B^\mu, 
\end{eqnarray}
where we omitted the vortical current due to technical difficulties. 
Thus, our calculations are applicable when vorticity (times chemical
potential) is not large
compared to the strength of the magnetic fields, which is the case in
heavy-ion collisions considered in this work.
In order to solve hydrodynamic equations, 
we also need to employ an appropriate equation of state,
the pressure as a function of the energy 
and charge densities $P = P (\varepsilon,n,n_5)$.
We discretize hydrodynamic equations above 
in the 
expanding coordinates $x^\mu = (\tau, x, y, \eta_s)$ 
and solved numerically, 
where $\tau = \sqrt{t^2 - z^2}$ is proper time, 
and $\eta_s = \frac{1}{2}\log \frac{t+z}{t-z}$ is space-time rapidity.  
This implementation allows us to study 
the dynamics of a fluid and anomalous transport phenomena 
simultaneously. 

\emph{Application to heavy-ion collisions.---}
It has been argued that experimental signal of macroscopic anomalous
transport effects can be obtained in heavy-ion collisions
\cite{Burnier:2011bf,Burnier:2012ae,Wang:2012qs,Ke:2012qb}.
Strong magnetic fields are expected to be produced especially in
off-central heavy-ion collisions, along which the current flows as a
result of axial anomaly.
However, the evidence of such transport effects has been elusive, 
mainly due to the lack of quantitative theoretical predictions. 
Numerical simulations of anomalous hydrodynamics enable us 
to gain an insight into the anomalous transport in the QGP.

For that purpose, 
we apply the anomalous hydrodynamic model 
introduced above to the space-time evolution of the QGP created in off-central 
heavy-ion collisions. 
Based on the results of the simulations, 
we discuss the implications in the experiments
We here consider a plasma composed of massless gluons and 
two flavor quarks (up/down) with corresponding electric charges.
As an equation of state, we employ that of the ideal gas with 
two-component massless fermion and gluons, 
$P(\varepsilon, n , n_5)= \varepsilon /3$~%
\footnote{
Let us note that the use of the ideal equation of state corresponds to a
conservative estimate.
This is because the velocity of the chiral magnetic wave is 
inversely proportional to the axial susceptibility of the plasma, 
and this velocity is minimal for the ideal equation of state. 
Since the magnitude of the velocity determines how fast the charges are 
transported, the estimate of the charge separation in terms of the 
ideal EOS is a conservative one. 
}.
This simplification does not hamper the 
essential features of the anomalous transport effects in the dynamics.

\begin{figure}[tbp]
 \begin{center}
  \includegraphics[width=70mm]{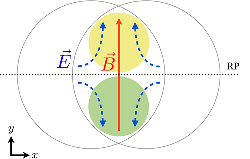}
 \end{center}
 \caption{
(Color online)
A typical configuration of electromagnetic field
in the transverse plane in non-central collisions.
The two circles indicate the edge of the colliding nuclei.
The solid line (red) shows the magnetic field and 
the dashed lines (blue) show the electric field. 
The inner product between electric and magnetic fields $\vec{E} \cdot \vec{B}$ 
becomes positive (negative) on the upper (lower) side of reaction plane (RP).
}
\label{fig:EMfield}
\end{figure}

Let us specify the initial values and time evolutions of electromagnetic
fields appropriate for the heavy-ion collisions.
A typical configuration of the electromagnetic field 
in off-central heavy-ion collisions is illustrated 
in Fig.~\ref{fig:EMfield}. 
The magnetic field has a special direction perpendicular to
the reaction plane (RP) and 
the electric field tends to direct out-of-plane.
We have defined the origin of the spatial coordinate as the middle of
the centers of the colliding nuclei and $x$ and $y$ axes 
as in Fig.~\ref{fig:EMfield}. 
We parametrize the evolution of background electromagnetic fields as 
\begin{align}
  eB_y (\bm{x}_\perp, \eta_s, \tau) 
  &=
  eB_0\frac{b}{2R} 
  \exp\left( - \frac{x^2}{2\sigma_{x}^2} 
    -  \frac{y^2}{2\sigma_{y}^2} -  \frac{\eta_s^2}{2\sigma_{\eta_s}^2}
    - \frac{\tau}{\tau_B} \right) ,  \\ %\nonumber \\ \\
  eE_y (\bm{x}_\perp, \eta_s, \tau)  
  &=  y \times eE_0\frac{b}{2R}  
  \exp\left( - \frac{x^2}{2\sigma_{x}^2} 
    -  \frac{y^2}{2\sigma_{y}^2} -  \frac{\eta_s^2}{2\sigma_{\eta_s}^2}
    - \frac{\tau}{\tau_E} \right) ,  %\nonumber \\
\end{align}
where $\sigma_x,~\sigma_y,~\sigma_{\eta_s}$ denote 
the spatial widths of the fields for each direction, 
$\tau_{B(E)}$ the duration time of the magnetic (electric) fields, 
$R = 6.38~\mathrm{fm}$ the radius of the a gold nucleus, 
and $b$ the impact parameter.
Other components of the electromagnetic fields 
are set to zero.
We set the widths as $\sigma_x = 0.8 (R - b/2),~ \sigma_y = 0.8\sqrt{R^2 - (b/2)^2}$, 
and $\sigma_{\eta_s} = \sqrt{2}$,
so that the fields are applied to the region the matter exists. 
The strength of the fields are taken to be proportional to the impact
parameter \cite{Kharzeev:2007jp}. 
The magnitudes of the electric and magnetic fields are set to 
$eB_0 = 0.5 \ \mathrm{GeV}^2,~ eE_0 = 0.1 \ \mathrm{GeV}^2$, 
which lead to $eB_y ( \bm{0}, 0, \tau_{\mathrm{in}}) \sim (3m_\pi)^2,
eE_y ( \bm{0}, 0, \tau_{\mathrm{in}}) \sim m_\pi^2$ 
and are consistent with the estimation in Refs.~\cite{Bzdak:2011yy, Deng:2012pc}.
The lifetimes \cite{McLerran:2013hla,Tuchin:2013apa} of electric and magnetic fields
are controlled by 
$\tau_B$ and $\tau_E$. 
The quantity $\vec{E} \cdot \vec{B}$ is
positive(negative) above(below) the reaction plane
(see also Fig.~\ref{fig:EMfield}), 
which leads to the generation of axial charges through the source term in Eq.~(\ref{eqn-accon}).

Let us describe the initial conditions for the fluid. 
The initial time for hydro is chosen to a standard value, $\tau_{\mathrm{in}}=0.6~\mathrm{fm}$. 
The temperature profile of the fluid is initialized using the modified BGK model
\cite{Brodsky:1977de, Adil:2005qn, Hirano:2005xf}, using parameters fitted to reproduce the
charged hadron multiplicities at RHIC energy. 
The initial axial charge is set to zero, 
$n_{5}(\bm{x}_\perp, \eta_s,\tau_{\mathrm{in}}) = 0$. 
In order to incorporate the ``stopping'' of the vector charges, 
we take the initial vector chemical potential as 
$\mu (\bm{x}_\perp,\eta_s,\tau_{\mathrm{in}}) 
= c T (\bm{x}_\perp,\eta_s,\tau_{\mathrm{in}})$,
where the parameter $c$ is a measure of the amount of the stopped vector charges. 
The value of $c$ is varied between 0 and 0.03, which results in the
variation of the charge asymmetry of the produced particles. 

\begin{figure}[tbp]
 \begin{center}
  \includegraphics[bb=100 60 295 239,clip,width=40mm]{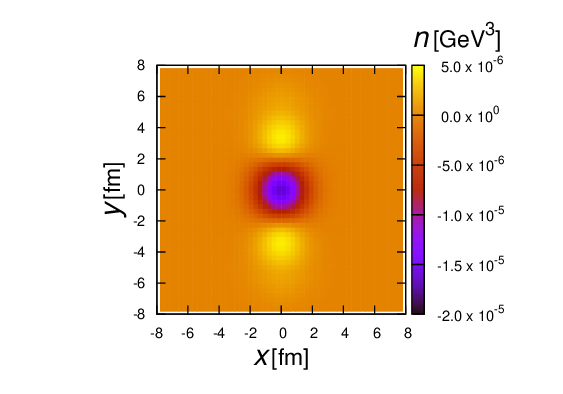}
  \includegraphics[bb=100 60 295 239,clip,width=40mm]{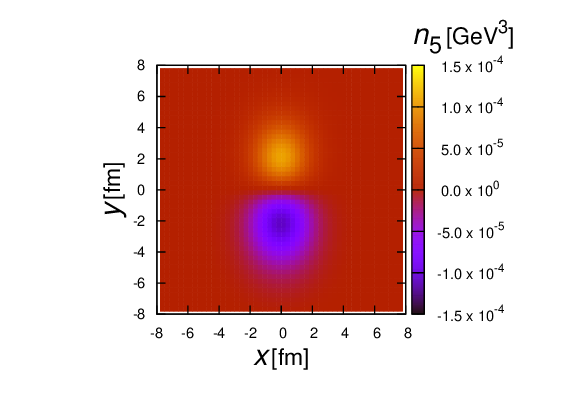}
 \end{center}
 \caption{
(Color online)
Vector (left) and axial (right) charge density distributions at 
$\tau=3~\mathrm{fm}$ 
in the $\eta_s=0$ plane, which is 
parallel to 
the external magnetic field. 
Because of the anomaly source and CME, 
vector (axial) charge has quadrupole (dipole) distribution 
even if initial vector charge density is zero. 
}
\label{fig:n-n5-t6-no-charge}
\end{figure}

Let us discuss the charge profiles developed by the anomalous
hydrodynamics.
Figure~\ref{fig:n-n5-t6-no-charge} shows the vector and axial charge 
densities in the transverse plane at mid-rapidity ($\eta_s = 0$)
at $\tau=3~\mathrm{fm}$. 
In this simulation, 
we choose $b=7.2~\mathrm{fm}$ (20-30$\%$ in centrality) and $c=0$, 
namely, both of the vector and axial charges are zero initially. 
The vector charges show quadrupole distributions (left), while
the axial charges show dipole-like distributions (right).
This is caused by the interplay between the axial charge generation through
anomaly equation (\ref{eqn-accon}) and 
anomalous transport effects \cite{Stephanov:2013tga}. 
First, 
the axial charge is generated through the term
$\vec{E} \cdot \vec{B}$ in Eq.~(\ref{eqn-accon}). 
This results in the dipole-like distributions of axial charge since 
the sign of $\vec{E} \cdot \vec{B}$ changes between above and below the
reaction plane (Fig.~\ref{fig:EMfield}).
Then, the vector current flows in the presence of axial charge
densities (CME), 
which makes the quadrupole distribution of vector charges \cite{Stephanov:2013tga}.
At the same time, 
the fluid expands radially because of the large pressure
gradients, and this would result in the quadrupole charge deformation in
the measured hadrons. 
The mechanism above is different from the 
formation of charge quadrupole from nonzero initial baryon densities, 
which correspond to lower energy collisions
\cite{Burnier:2011bf,Burnier:2012ae}. 
The above result indicates that, 
even though there is no vector charge initially, 
which corresponds to higher energy collisions,
a charge quadrupole deformation occurs \cite{Stephanov:2013tga}.
Our calculations demonstrate that, a charge quadrupole can be formed 
at high energy collisions as well as low energy ones, 
as a result of the anomalous transport effects and axial charge
generation through anomaly. 

\emph{Implications on experimental observables.---}
Here we discuss implications of the anomalous hydrodynamics
for the results in heavy-ion collisions experiments.
Since the initial shape of the QGP produced in off-central collisions is
like an almond (see Fig.~\ref{fig:EMfield}), the pressure gradients lead to 
the anisotropic momentum distribution of emitted particles.
The elliptic flow $v_2$ is an experimental signal of this
anisotropic momentum distribution \cite{Ollitrault:1992bk}.
If a charge quadrupole is formed during the QGP evolution, 
charge-dependent particle distributions in the final state should also
be affected. 
As a result, the elliptic flow of negatively-charged particles 
is expected to be larger than 
that of positively-charged particles: $v_2^- > v_2^+$. 

In order to obtain the momentum distribution of particles, 
we employ the Cooper-Frye formula \cite{Cooper:1974mv} 
on an isothermal freezeout hypersurface with
$T_{\mathrm{fo}}=160~\mathrm{MeV}$, 
\begin{equation}
 p^0\frac{dN}{d^3 p} = \frac{d}{(2\pi)^3} \int_{\Sigma_{T_{\mathrm{fo}}}}
  \frac{p^\mu d\sigma_\mu}{\exp [\left( p^\mu u_\mu - \mu \right)/T_{\mathrm{fo}}] \mp 1}.
\label{eq-CF}
\end{equation}
Here $d\sigma_\mu$ is a hypersurface element, 
and $d$ the number of degree of freedom and $\mp$ corresponds to Bose/Fermi 
distribution for the produced particles.
By applying the Cooper-Frye formula (\ref{eq-CF}), 
we can calculate charge-dependent azimuthal particle distributions,
 \begin{eqnarray}
\frac{ d N_\pm }{d \phi} (\phi) &=& \bar N_\pm
\left( 1 + 2 \sum_n v_n^\pm \cos n\phi \right), 
 \end{eqnarray}
where the  azimuthal angle $\phi$ is measured from the $x$-axis and 
$\bar N_\pm$ is defined as the angle average of the number distribution,
\begin{equation}
 \bar N_\pm \equiv \int \frac{d\phi}{2\pi}
\frac{dN_\pm}{ d \phi}.
\end{equation}
Then we can obtain the charge-dependent elliptic flow parameter by 
$v_2^\pm = \langle \cos (2 \phi)  \rangle_\pm$,
where the bracket $\langle ... \rangle_\pm$ denotes averages over the 
azimuthal particle distributions for positively (negatively) charged
particles. 
In the current work, 
considering an experimental setup for STAR Collaboration in RHIC \cite{Wang:2012qs,Ke:2012qb}, 
we use charged particles within $0.15 <p_T <12~\mathrm{GeV}$, 
and$|\eta|<1$ to estimate the net charge asymmetry 
$A_{\pm} \equiv (\bar{N}_+ - \bar{N}_-)/(\bar{N}_+ + \bar{N}_-)$,
and within $0.15 <p_T <0.5~\mathrm{GeV}$, and  
$|\eta| <1$ to calculate the charged elliptic flow $v_2^\pm$.
But we do not consider the event-by-event fluctuation 
effects coming from acceptance limit of the detector 
\footnote{
We take $c$ 
to be positive, so $A_\pm$ is always positive.
In real experiments, negative $A_\pm$ is allowed due to the limited
acceptance \cite{Bzdak:2013yla}. 
}
.

We calculate $\Delta v_2^\pm$ as a function of  $A_\pm$ through 
different initial charge distributions by changing $c$
as a parameter which controls the initial charge fluctuation.
$\Delta v_2^\pm(A_\pm)$ is a key quantity in the discussion
of the signals of anomalous transports.
Recently, the characteristic linear dependence of $\Delta v_2^\pm$ on $A_\pm$, 
$\Delta v_2^\pm = 2 r_e A_\pm + O (A_\pm
^2)$, 
is argued to be one of the possible signals of the anomalous
transport \cite{Burnier:2011bf,Burnier:2012ae}. 
The experimental data also show the same tendency
\cite{Wang:2012qs,Ke:2012qb}. 
However, according to the simulation of anomalous hydrodynamics, 
we conclude that the linear tendency cannot be regarded as a 
direct consequence of
anomalous transports, at least for $A_\pm >0$, 
since it is present even when we switch off the anomalous effects 
(namely, taking $C=0$ in the EOMs)
\footnote{
In Ref.~\cite{Bzdak:2013yla}, 
it is pointed out that 
the linear tendency can be reproduced by an event-by-event hydrodynamic model 
which incorporates the limited acceptance of detectors,
local charge conservation, and rapidity dependence of $v_2$.
This is another mechanism to produce linear dependence of $\Delta v_2^\pm$
on $A_{\pm}$.
}.

\begin{figure}[tbp]
 \begin{center}
  \includegraphics[width=70mm]{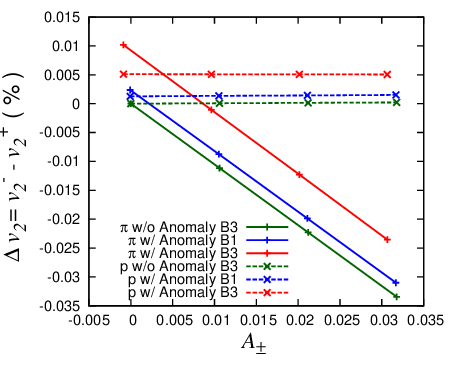}
 \end{center}
 \caption{
(Color online)
$\Delta v_2^\pm \equiv v_2^- - v_2^+$ is shown as a function of 
the net charge asymmetry $A_\pm$ 
(Solid lines for $\pi^\pm$ and dashed ones for $p\bar{p}$).
Green lines are the results in the absence of the anomaly ($C=0$).
Blue and red lines correspond to the cases with anomaly, 
where the parameters are set to 
$(\tau_B, \tau_E)= ( 1.0 \ \mathrm{fm}, 1.0 \ \mathrm{fm} )$ and 
$(\tau_B, \tau_E)= ( 3.0 \ \mathrm{fm}, 1.0 \ \mathrm{fm} )$, respectively.
Each dot corresponds to the different 
initial chemical potential 
$c$ 
from $0.0$ to $0.03$ with $\mu(\tau_{\mathrm{in}}) = c T (\tau_{\mathrm{in}})$.
All lines show the linear relation between $\Delta v_2^\pm$ and $A_\pm$
 with close to the same slopes, while the values of 
 $\Delta v_2^\pm(A_\pm=0)$  are different. 
}
\label{fig:A-dv_2}
\end{figure}

Figure \ref{fig:A-dv_2} shows $\Delta v_2^\pm$ as a function of  the
net charge asymmetry $A_\pm$. 
Different lines correspond to different conditions. 
Here solid lines show results for pions: $v_2(\pi^-) -v_2(\pi^+)$, 
whereas dashed lines do for protons: $v_2(\bar{p}) - v_2(p)$. 
Green lines are the results in the absence of the anomaly:
the anomaly constant $C$ is set to zero
and the CME/CSE do not take place.
Blue and red lines correspond to the cases with the anomaly, 
where the parameters $(\tau_B, \tau_E)$ are set to 
$( 1.0 \ \mathrm{fm}, 1.0 \ \mathrm{fm} )$ and 
$ ( 3.0 \ \mathrm{fm}, 1.0 \ \mathrm{fm} )$, respectively.
We find a linear relation between 
$\Delta v_2^\pm$ and $A_\pm$ for all cases.

We first note that the intercepts of these lines, $\Delta v_2^\pm(A_\pm = 0)$, 
are sensitive to anomalous effects robustly. 
While $\Delta v_2^\pm (0) = 0$  
in the absence of anomaly ($C=0$), 
it is finite and positive in the presence
of anomaly, as shown in Fig.~\ref{fig:A-dv_2}.
Since the magnitude of the intercepts decreases significantly 
when we turn off the electric field ($eE_{\mathrm{max}} = 0 $), 
the finite intercepts are caused by the interplay of the 
axial charge generation by $\vec{E} \cdot \vec{B}$ term and subsequent
anomalous charge transport. 
The magnitude of intercepts also depends on the duration time of magnetic fields:
larger duration time results in larger $\Delta v_2^\pm(0)$.
This fact can be useful in inferring the dynamics of electromagnetic
fields in the plasma. 
However, the value of the slope parameter $r_e$ might not be a good signal of
anomaly-induced transports, because it also depends on the choice 
of the quantum statistics of the emitted particles, namely, the sign in the
Cooper-Frye integral (\ref{eq-CF}).
It can be nonzero even if the anomalous transport effects are absent. 
If observed particles are fermions (bosons), 
the slope $r_e$ becomes positive (negative), 
while $r_e \simeq 0$ in the case with particles obeying the classical 
statistics 
\footnote{
If we employ the classical statistics, namely the Maxwell-Boltzmann statistics,
we obtain exactly zero (small but nonzero) slope in the absence (presence) of anomaly, 
which is consistent with results in \cite{Yee:2013cya}.
}.
Since the proton mass is significantly larger than the freezeout
temperature, protons approximately obey the classical statistics and
the slope is quite small. 
According to the current simulation results, 
the values of the slope parameters are mainly driven by
the quantum statistics of emitted particles,
rather than anomaly transport effects.
Let us also note that the contribution from  the quantum
statistics to the slope parameter vanishes in a certain limit,
in which produced particles are massless and
$p_T$ integral is performed from $0$ to $\infty$ \cite{Hatta-Xiao}.
Thus, the contribution might be due to the 
combined effects of quantum statistics, finite $p_T$ cuts, and finite
mass.
Existence of other conserved charges would also affect the values of charge
dependent elliptic flow for different hadron species. 
Incorporating those factors requires much more systematic study,
that should be an important future work.

Let us comment on the uncertainties 
in the current calculations. 
Firstly, 
we have neglected the conductive current
\cite{Deng:2012pc,Stephanov:2013tga, Hirono:2012rt} in this study. 
It also makes a contribution to 
charge quadrupole deformation in the QGP (see Fig.~\ref{fig:EMfield}).
Secondly, we have also neglected the back reaction of 
matter to electromagnetic fields.
The back reactions would 
affect the duration time of the magnetic field, 
which is treated as parameters in this study. 
In order to quantify this effect, we have to calculate the time
evolutions of dynamical electromagnetic fields simultaneously. 
Thirdly, the effects of the QCD sphalerons can be also important since 
it reduces the axial charge.
Consideration of these effects is left for future works.

\emph{Summary.---}
In this Letter, we have performed the first numerical simulation of 
anomalous hydrodynamics.
We have developed a numerical code to study anomalous transport
effects quantitatively and applied the code to heavy-ion collision experiments.

We have demonstrated
the chiral magnetic wave in the expanding quark-gluon plasma 
and found that the charge quadrupole deformation is produced even if
there is no initial vector charge. 
We have also shown that 
the finite slope parameter $r_e$
is not a proper signal for
the anomalous transport effects, 
since the slope can be also present even
in the absence of anomaly.
The results of simulations indicate that 
the intercepts, $\Delta v_2^\pm(A_\pm=0)$, are robustly sensitive to the
anomalous effects. 

\section*{Acknowledgement}
The authors thank 
Y.~Hatta, B.~W.~Xiao, A.~Bzdak, K.~Fukushima, K.~Murase, R.~Kurita and
Y.~Tachibana for useful discussions.
M.~H. and Y.~H are partially supported by JSPS Strategic Young Researcher
 Overseas Visits Program for Accelerating Brain Circulation (No.R2411).
M.~H was supported by RIKEN Junior Research Associate Program and 
the Special Postdoctoral Researchers Program at RIKEN. 
Y.~H. was partially supported by the Japan Society for the Promotion of Science for
Young Scientists. 
The work of T.~H. is supported by
Grant-in-Aid for Scientific Research
No.~25400269.

%%%%%%%%%%%%%%%%%%%%%%%%%%%%%%

\end{document}